\documentclass[pra,10pt,twocolumn,superscriptaddress,floatfix,showpacs]{revtex4-1}

\usepackage{subfig}
\usepackage{graphicx} % Package to insert exteral figures
\usepackage{epstopdf}
\usepackage[utf8]{inputenc}
\usepackage[T1]{fontenc}     %Output what you want e.g., é, l, a, ü
\usepackage[british]{babel}  %Do hyphenation according to british english
\usepackage[sc,osf]{mathpazo}
\usepackage[scaled=0.86]{berasans}  % URL font that go well wtih palatino
\usepackage[colorlinks=true, citecolor=blue, urlcolor=blue]{hyperref}  %Hyperlinks (pink, green, blue)
\usepackage[babel]{microtype}  %Improves text justification
\usepackage{amsmath,amssymb,amsthm,bm,amsfonts,mathrsfs,bbm} %Usefull math packages

\usepackage{xspace}  %Useful to add space in macros
\usepackage{pgfplots}
\usepackage{xcolor,colortbl}
\def\ba{\begin{equation}}
\def\ea{\end{equation}}
\def\bea{\begin{eqnarray}}
\def\eea{\end{eqnarray}}
\def\ben{\begin{equation*}}
\def\een{\end{equation*}}
\def\bean{\begin{eqnarray*}}
\def\eean{\end{eqnarray*}}
\def\bma{\begin{mathletters}}
\def\ema{\end{mathletters}}
\def\bi{\begin{itemize}}
\def\ei{\end{itemize}}

\newcommand{\be}{\begin{equation}}
\newcommand{\ee}{\end{equation}}

\newcommand{\kommentar}[1]{}

\newcommand{\forget}[1]{}

\begin{document}

\title{Conditional steering under von-Neumann scenario}
%#####################################

\author{Kaushiki Mukherjee}
\email{kaushiki_mukherjee@rediffmail.com}
\affiliation{Department of Mathematics, Government Girls’ General Degree College, Ekbalpore, Kolkata-700023, India.}

\author{Biswajit Paul}
\email{biswajitpaul4@gmail.com}
\affiliation{Department of Mathematics, South Malda College, Malda, West Bengal, India}

\author{Sumana Karmakar}
\email{sumanakarmakar88@gmail.com}
\affiliation{Department of Applied Mathematics, University of Calcutta, 92, A.P.C. Road, Kolkata-700009, India.}

\author{Debasis  Sarkar}
\email{dsarkar1x@gmail.com}
\affiliation{Department of Applied Mathematics, University of Calcutta, 92, A.P.C. Road, Kolkata-700009, India.}

\author{Amit Mukherjee}
\email{amitisiphys@gmail.com}
\affiliation{Physics and Applied Mathematics Unit, Indian Statistical Institute, 203,B. T. Road, Kolkata 700108 , India.}

\author{Some Sankar Bhattacharya}
\email{somesankar@gmail.com}
\affiliation{Physics and Applied Mathematics Unit, Indian Statistical Institute, 203,B. T. Road, Kolkata 700108 , India.}

\author{Arup Roy}
\email{arup145.roy@gmail.com}
\affiliation{Physics and Applied Mathematics Unit, Indian Statistical Institute, 203,B. T. Road, Kolkata 700108 , India.}

%########################################

\begin{abstract}
In [\href{http://www.sciencedirect.com/science/article/pii/037596019290711T}{Phys. Lett. A {\bf 166}, 293 (1992)}] Popescu characterized nonlocality of pure $n$-partite entangled systems by studying bipartite violation of local realism when $n-2$ number of parties perform projective measurements on their particles. A pertinent question in this scenario is whether similar characterization is possible for $n$-partite mixed entangled states also. In the present work we have followed an analogous approach so as to explore whether given a tripartite mixed entangled state the conditional bipartite states obtained by performing projective measurement on the third party, demonstrate a weaker form of nonlocality, quantum steering. We also compare this new phenomenon of conditional steering with existing notions of tripartite correlations. Interestingly the tripartite state need not be genuinely entangled to demonstrate conditional steering.
\end{abstract}

%\pacs{03.65.Ud, 02.50.Le, 03.67.Ac}

\maketitle

%########################################
	
\section{Introduction}
In recent past quantum inseparability, mostly known as entanglement \cite{Reviw Ent} has been proven to be an important ingredient for various information processing tasks viz. teleportation \cite{Bennett92}, dense coding \cite{Bennett93}, where this non-classical resource outperforms any classical resource. Beside \emph{entanglement} quantum mechanics possesses more such striking phenomena where joint probability distributions arising from composite systems can not be reproduced by any local realistic theory as demonstrated by the pioneering work of John Bell known as \emph{Bell's Theorem} \cite{Bell64}. Non-local quantum correlations have been found to be fundamental resource for device independent certification of randomness \cite{random}, cryptography\cite{key}, dimension witness \cite{dw}, Bayesian game theoretic applications \cite{game}, quantum network\cite{km1} etc. In such cases the existence of entanglement only is not sufficient to guarantee the advantage. Recently there have been considerable amount of study regarding another non-classical phenomena commonly known as quantum steering \cite{Wiseman07}. The concept of steering was originally perceived by Schr\"{o}dinger. It describes a scenario where one of the subsystems controls the conditional state of another spatially separated party by sharing a suitable quantum state and applying appropriate measurements on his or her part. Quantum steering has been explained as a semi device independent entanglement witness in contrary to the Bell non-local device independent scenario. In that sense steering can be thought as a weaker form of non-locality. Thus entanglement, quantum steering and Bell-nonlocality can be seen as a hierarchy of quantum correlations, where all entangled states may not be steerable and nonlocal, all steerable states are necessarily entangled but not necessarily nonlocal and all non-local states are both entangled and steerable.

The idea of entanglement, nonlocality and steering in more than two parties is not as simple and straight forward as the bi-partite case. On one hand complexity in analyzing correlations increases with number of parties, on the other hand multipartite correlations consist of completely new features absent in the bi-partite scenario. Thus multipartite extension of correlations has gained researchers' attention in recent times\cite{ban,kmb}. Quantum correlation among spatially-separated parties can be visualized as a network with a precise global description lacking a sum of local descriptions interpretation. In this context, it is of immense importance if one can comment on the correlation present in the global system by investigating the conditional bipartite marginals only.

Extending the previous results \cite{Gisin91, Gisin92} Popescu demonstrated the nonlocality of $n$ qubit pure entangled states by using only bipartite Bell-CHSH inequalities\cite{Popescu92}. He in particular argued that for any $n$ qubit pure entangled state, if there exists a global projective measurement on any possible $(n-2)$ systems out of the $n$ systems that leaves the remaining conditional bipartite state as a pure entangled state, resulting in a violation of bipartite Bell-CHSH inequality, then that guarantees nonlocality of the pure $n$ qubit entangled state. This argument therefore characterizes nonlocality of a pure $n$ partite entangled state. However, from experimental perspectives, it will be more interesting to pose the analogous question regarding the characterization of nonlocality of $n$-partite mixed entangled states. In the present work we intend to contribute in this direction. To be more precise, our present topic of discussion mainly explores the possibility of revealing nonlocality of any mixed entangled tripartite state by observing nonlocality of any conditional bipartite state obtained from the tripartite state due to suitable projective measurement performed by any of the three parties. For that we define a new form of tripartite nonlocality in which the tripartite state is said to be nonlocal if any of the conditional bipartite states demonstrates steering nonlocality.

The rest of the article is ordered as follows: in Sec.\ref{mot} we discuss the motivation that further underlines importance of this work. Sec.\ref{prelim} provides the necessary preliminaries to set the premise to present our results. In Sec.\ref{cs} we define the concept of conditional steering and present our findings. We conclude in Sec.\ref{concl}.

\section{motivation}\label{mot}
Nonlocal correlations obtained from entangled quantum states are useful for various information processing tasks\cite{random}. In this context one can try to capture the ability of a state, which is `local' based on all the existing notions of nonlocality, to reveal a weaker form of nonlocality. For instance, in \cite{manik}, the author showed that mixed entangled bipartite quantum states having a local description can be used to certify randomness in a measurement device independent randomness-certification task.

From practical point of view, the topic becomes more interesting if one tries to characterize nonlocal behavior of multipartite mixed entangled quantum states. This basically motivates our present discussion. Precisely speaking, we define a new form of tripartite nonlocality in the same spirit of Popescu(discussed before). In contrast to Popescu's approach\cite{Popescu92}, we have considered steering nonlocality, a weaker form of nonlocal correlation of the conditional states arising due to some suitable projective measurements by one of the three parties sharing the tripartite state. This in turn emerges as a useful tool to  characterize nonlocality of some mixed tripartite states which are local in some specific Bell scenario\cite{Sliwa03} and also may not even be steerable\cite{cavals}. This fact enhances the possibility of the corresponding tripartite state to be used as a suitable candidate in some information theoretic task. Interestingly we observe that for any tripartite state to be nonlocal according to our notion of nonlocality, it does not need to be genuinely entangled. So any biseparable state can also exhibit this form of nonlocality. Besides, we also explore other related issues thereby encountering some additional implications in this context.

Having sketched our motivation, we are now in a position to present our work. But first we provide with some mathematical tools which are to be used later in our discussion.

\section{Preliminaries}\label{prelim}
\subsection{Bipartite Steering(EPR steering)}  In a bipartite scenario the correlation terms are steerable from Alice to Bob if they are inexplicable by any local hidden variable(LHV) -local hidden state(LHS) model \cite{Wiseman07}:
\begin{equation}\label{pst1}
P(a, b|x, y)=\sum_{\lambda} p_{\lambda}P(a|x, \lambda)\textmd{Tr}[\widehat{\Pi}(b|y)\rho_2(\lambda)]
\end{equation}
where $\lambda$ denotes the hidden variable, $a$, $b$ denote local results, $x,y$, the corresponding local inputs and $\widehat{\Pi}(b|y)$ is the projection operator corresponding to an observable
characterized  by Bob's setting $y$, associated with the eigenvalue $b$ and $\rho_2(\lambda)$ corresponds to some pure state of Bob’s system, parameterized by hidden variable $\lambda$. So if $\rho_{AB}$ be  a bipartite quantum state shared in between Alice and Bob. If the correlations arising due to their measurements, lack a decomposition in the above form(Eq.(\ref{pst1})) then $\rho_{AB}$ is said to be steerable from Alice to Bob. Steering being an asymmetric phenomenon, $\rho_{AB}$'s steerability from  Alice to Bob does not necessarily imply its steerability from Bob to Alice\cite{Bowles14}.
\subsection{Steering Criteria} Here we discuss two existing criteria for detecting bipartite steering \cite{Zukowski,Jevtic15}.
\subsubsection{$S_1$}
In \cite{Zukowski} Zukowski et.al. gave a criterion which suffices to detect steerability of a bipartite quantum state. Any bipartite qubit density matrix can be decomposed as:
\begin{equation}\label{st4}
\rho_{AB}=\frac{1}{2^2}\sum_{i_1,i_2=0}^{3}t_{i_1i_2}\sigma^1_{i_1}\bigotimes\sigma^2_{i_2}
\end{equation}
where $\sigma^k_0,$ denotes the identity operator in the Hilbert space of qubit k and $\sigma^k_{i_k},$ are the Pauli operators along three perpendicular directions, $i_k=1,2,3$.
 The components $t_{i_1i_2}$ are real and given by:
 \begin{equation}\label{st4i}
 t_{i_1i_2}=\textmd{Tr}[\rho_{AB}\sigma^1_{i_1}\bigotimes\sigma^2_{i_2}],\,i_1,i_2\in\{0,1,2,3\}.
 \end{equation}
 Using basic geometric approach in \cite{Zukowski}, Zukowski et.al. showed that the quantum state($\rho_{AB}$) shared between Alice and Bob is steerable(from Alice to Bob) if the corresponding correlation functions satisfy the following inequality:
\begin{equation}\label{st8}
||t||_{\infty}<\frac{2}{3}||t||^2
\end{equation}
where $||t_{\infty}||$ and $||t||$ denote spectral norm and Hilbert-Schmidt norm  of $t$ respectively. Any quantum state which satisfies this criteria(Eq.(\ref{st8})) is steerable in nature. We define a measure of bipartite steering $S_1$ where:
\begin{equation}\label{st7i}
S_1\equiv  ||t||_{\infty}-\frac{2}{3}||t||^2
\end{equation}
Hence any negative value of $S_1$ indicates that the bipartite state($\rho_{AB}$) shared between Alice and Bob is steerable from Alice to Bob.
However, if any state fails to satisfy it(Eq.(\ref{st8})), i.e., for any positive value of $S_1$, the steerability of the state cannot be guaranteed.
\subsubsection{$S_2$}
Given density matrix representation(Eq.(\ref{st4})) of a bipartite state($\rho_{AB}$), the correlation tensor matrix $t$ having elements $t_{ij}(i,j\in\{1,2,3\})$ given by (Eq.(\ref{st4i})) can always be represented in the form: $t=R^ADR^B$, where $D$ is a diagonal matrix $\textmd{diag}(\mathcal{T}_1,\mathcal{T}_2,\mathcal{T}_3)$ and $R^A, R^B$ stand for proper rotations($\in SO_3$). Clearly, if $t$ itself is a diagonal matrix then $\mathcal{T}_i=t_{ii}(i\in\{1,2,3\})$. Let for any $i\in\{1,2,3\}$ $A_i$ and $B_i$ denote the modified versions of the terms $t_{i0}$ and $t_{0i}$ respectively after applying $R^A$ and $R^B$ on first and second qubits respectively. As any rotation($R^A,\,R^B$) can be implemented by local unitary operations which correspond to local basis change, so any property of the state($\rho_{AB}$) remains intact. Hence correlation matrix of any state($\rho_{AB}$) can be diagonalized without altering steerability of $\rho_{AB}$.\\
Let $\rho_{AB}$ be shared between Alice and Bob. Let Bob can measure anyone of $\mathbb{I}_2\bigotimes \sigma_3$ or $\mathbb{I}_2\bigotimes \sigma_\mu$, with $\sigma_\mu=\sigma_1 \cos\mu+\sigma_2\sin\mu$($\mu$$\in$$[0,2\pi]$) and Alice can measure on her qubit Hermitian observables of the form $A_1\bigotimes\mathbb{I}_2$, $A_\mu=\sigma_\mu\bigotimes\mathbb{I}_2.$($\mathbb{I}_{2}$ denotes $2\times 2$ identity matrix) By choosing $A_1=\sigma_3$ and $A_\mu=\sigma_1 \textmd{sign}(\mathcal{T}_1)\cos\mu+\sigma_2 \textmd{sign}(\mathcal{T}_2)\sin\mu$, Zevtic \textit{e.t al.}\cite{Jevtic15} modified a pre-existing EPR steering inequality\cite{Jevtic16}, thereby framing the nonlinear EPR steering inequality:
\begin{equation}\label{st8i}
	\begin{split}
	 |\mathcal{T}_1|+|\mathcal{T}_2|-\frac{2}{\pi}(\sqrt{(1+A_3)^2-(\mathcal{T}_3+B_3)^2}\\+\sqrt{(1-A_3)^2-(\mathcal{T}_3-B_3)^2})\leq 0
	\end{split}
\end{equation}
If correlations arising due to local measurements on any bipartite state $\rho_{AB}$ can be explained by a LHV-LHS model(Eq.(\ref{pst1})), then $\rho_{AB}$ necessarily satisfies the above inequality(Eq.(\ref{st8i})). So violation of this inequality is sufficient to claim that $\rho_{AB}$ is steerable from Alice to Bob. Considering all possible permutations of $\mathcal{T}_i,\,A_i,\,B_i,i\in\{1,2,3\}$, the sufficient criteria for detecting steerability of $\rho_{AB}$ was ultimately framed as\cite{Jevtic15}:
\begin{equation}\label{st8ii}
\textmd{max}\{h(i,j,k),h(j,k,i),h(k,i,j)\}>0
\end{equation}
where
\begin{equation}\label{st8iii}
\begin{split}
h(i,j,k)= |\mathcal{T}_i|+|\mathcal{T}_j|-\frac{2}{\pi}(\sqrt{(1+A_k)^2-(\mathcal{T}_k+B_k)^2}\\+\sqrt{(1-A_k)^2-(\mathcal{T}_k-B_k)^2})
\end{split}
\end{equation}
We define:
\begin{equation}\label{st8iv}
S_2\equiv\textmd{max}\{h(i,j,k),h(j,k,i),h(k,i,j)\}
\end{equation}
as the second measure of bipartite steering. So any positive value of $S_2$ guarantees that the bipartite state($\rho_{AB}$) is steerable from Alice to Bob. \\
Having discussed about bipartite steering, we now deal with steerability of tripartite state: tripartite steering nonlocality\cite{cavals} and then genuine steering nonlocality\cite{Jeba}.
\subsubsection{Tripartite Steering Nonlocality}
Let $\rho_{ABC}$ be the state that Alice, Bob and Charlie share. A referee, who trusts the measurement device of only one party, say Alice, wants to check whether the correlations arising from $\rho_{ABC}$ are steerable or not. He will be sure that the tripartite correlations are steerable(from  Bob and Charlie to Alice) if the correlations lack a decomposition of the form\cite{cavals}:
\begin{equation}\label{steer1}
P(a,b,c|x,y,z)=\sum_{\lambda} q_{\lambda}\textmd{Tr}[\widehat{\Pi}(a|x)\rho_{A}^{\lambda}]P(b|y,\lambda)P(c|z,\lambda).
\end{equation}
 Here $\lambda$ is the hidden variable, $y,z$ and $b,c$ denote local inputs and outputs of Bob and Charlie respectively. $\widehat{\Pi}(a|x)$ denotes the projection operator corresponding to observable characterized  by Alice's setting $x$, associated with the eigenvalue $a$. $\rho_{A}^{\lambda}$ stands for some pure state of Alice's system which is classically correlated to the hidden variable $\lambda$. If the tripartite correlations are steerable, i.e., cannot be expressed in the above form(Eq.(\ref{steer1})), then the corresponding state $\rho_{ABC}$ is said to be \textit{steerable} from Bob and Charlie to Alice. In \cite{cavals}, an inequality (Eq.22 in \cite{cavals}) (based on the correlations generated) under some specific measurement settings(von-Neumann equatorial measurements) of the three parties (where the referee trusts only one party) was given, so as to detect steerability of the tripartite state involved. Violation of this inequality is only a sufficient criterion for detecting steerability of $\rho_{ABC}$ from Bob and Charlie to Alice. Hence if it is satisfied then nothing can be concluded about the steerability of $\rho_{ABC}$.\\
\subsubsection{Genuine Steering Nonlocality}
Now we are going to discuss the criteria of detecting genuine steering which have been explored recently\cite{Jeba}. Tripartite correlations $P(a,b,c|x,y,z)$ are said to be genuinely steerable\cite{Jeba} from one party, say Charlie to other two parties, say Alice and Bob, if those cannot be decomposed in the following form:

\begin{eqnarray}\label{jeva1}
   P(a,b,c&|&x,y,z)= \sum_{\lambda}q_\lambda \textmd{Tr}[\widehat{\Pi}(a,b|x,y)\rho_{AB}(\lambda)]P(c|z,\lambda)\nonumber\\
   &+&\sum_{\lambda}p_{\lambda}\textmd{Tr}[\widehat{\Pi}(a|x)\rho_{A}]Tr[ \widehat{\Pi}(b|y)\rho_{B}(\lambda)]P(c|z,\lambda)\nonumber\\
\end{eqnarray}
where Alice and Bob have access to qubit measurements whereas Charlie performs uncharacterized measurement.
  In \cite{Jeba}, the author gave an inequality to detect genuine tripartite steering(Svetlichny steering)based on Svetlichny inequality\cite{SVE}. The detection criterion is given as:
\begin{equation}\label{pst5}
\langle CHSH_{AB}z_1+CHSH^{'}_{AB}z_0\rangle_{2\times2\times ?}^{NLHS}\leq 2\sqrt{2}
\end{equation}
where $CHSH_{AB}$ and $CHSH_{AB}^{'}$ denote two inequivalent facets of the Bell-CHSH polytope. Here $NLHS$ denotes nonlocal hidden state and $2\times2\times ?$ implies that two parties Alice and Bob have access to qubit measurements whereas the third party Charlie does not trust his measurement device. The measurement settings of Alice and Bob should be orthonormal. If a given quantum state be such that the correlations arising due to measurements on it violate this inequality(Eq.(\ref{pst5})), then the state is genuinely steerable from Charlie to Alice and Bob. Similar steerability criteria from Alice to Charlie and Bob and from Bob to Charlie and Alice can be defined. So in totality if any state violates any one of these three criteria, then the state is genuinely steerable from one party to the remaining two parties. However if it does not violate any of these inequalities(sufficient to detect genuine steerability) then the state may not be genuinely steerable.
Having discussed about two measures($S_1$, $S_2$ of bipartite steering) along with the criteria of steering, we further mention a stronger notion of nonlocality in the next sub-section. We do this since our work introduces the new concept of conditional steering which is weaker compared to the existing notions of tripartite steering and nonlocality.
\subsection{Tripartite Nonlocality}
Let  Alice, Bob and Charlie, sharing state $\rho_{ABC}$, perform local measurements on their respective subsystems. The resulting correlations are nonlocal if those cannot be decomposed in the form:
\begin{equation}\label{pst2}
P(a,b,c|x,y,z)=\sum_{\lambda}p_{\lambda}P(a|x, \lambda)P(b|y,\lambda)]P(c|z,\lambda).
\end{equation}
and the corresponding state is said to be nonlocal. Such type of correlations are capable of  violating  tri-partite Bell inequality. For two input-two output scenario for each of the three parties, Sliwa defined a Bell-local polytope having $46$ facets\cite{Sliwa03}. The set of these $46$ facet inequalities of this polytope serves as a necessary and sufficient condition for detecting tripartite nonlocality.
\subsection{Negativity of entanglement}
Negativity of entanglement was  first introduced by Zyczkowski \textit{et. al} \cite{Zyczkowski} and then Vidal \textit{e.t al.} \cite{vidal} introduced it as a bipartite measure of  entanglement. An important property of this  measure is that, it is computable even for mixed states. Let $\rho_{ab}$  be a bipartite state of a composite system of dimension $d\otimes d'$. Then negativity of $\rho_{ab}$  is denoted by $\mathcal{N}_{a|b}(\rho_{ab})$ and is defined as \cite{vidal}
\begin{equation}\label{negativity}
\mathcal{N}_{a|b}(\rho_{ab})=\frac{||\rho_{ab}^{T_a}||_1-1}{d-1}
\end{equation}
where $\rho_{ab}^{T_a}$ denotes the partial transpose of $\rho_{ab}$ on the sub system $a$ and $||X||_1$ denotes the trace norm, i.e, $||X||_1=\textmd{Tr}(\sqrt{XX^{\dagger}})$. According to the definition , the quantity $\mathcal{N}_{a|b}(\rho_{ab})$ is equals to the twice sum of the absolute values of the negative eigenvalues of $\rho_{ab}^{T_a}$.\\

Negativity of entanglement is strongly related to the GHZ distillability of three-qubit system \cite{GHZ distillation}. A three-qubit state is GHZ distillable if negativity of each bipartite cut of the system is positive. So this measure of entanglement will be helpful to discuss our work from some information theoretic aspect. \\
In the next section we define our premise.

\section{Conditional Steering Nonlocality}\label{cs}
Let a tripartite state $\rho_{ABC}$ be shared between Alice, Bob and Charlie. Let Charlie first performs a projective measurement $\Pi_z^c$ on his qubit and broadcasts the output to rest. The resulting conditional bipartite state between the rest of the parties, i.e., Alice and Bob is denoted by,
\begin{equation}
\rho^{AB}_{\Pi_z^{c}}=\textmd{Tr}_c[\mathbb{I}_2\otimes\mathbb{I}_2\otimes\Pi_z^c~\rho_{ABC}].
\end{equation}
In a similar way, when Alice or Bob performs projective measurements on their qubits, corresponding conditional states
between the rest of the parties can be written as $\rho^{BC}_{\Pi_x^a}$ or $\rho^{AC}_{\Pi_y^b}$ respectively. Now we
want to check whether  $\rho^{AB}_{\Pi_z^c}$ is steerable from Alice to Bob. This in turn guarantees that Alice can
steer particle of  atleast one of the two remaining parties. Such type of tripartite states from which one can obtain
conditional states that are steerable, are defined as \textit{conditionally steerable} states(steerable from Alice to
Bob in this case) and corresponding tripartite correlation terms as\textit{ conditional steering} correlations. The
nomenclature is justified based on the fact that one of the three parties can steer qubit of one of the two remaining
 parties in spirit of Popescu \cite{Popescu92} where they ask the question for non local scenario. So any conditionally
 steerable tripartite state exhibits a weaker form of nonlocality which can be defined as \textit{conditional steering nonlocality}.
\subsection{Detecting conditional steering nonlocality}
\begin{figure}[htb]
	\centering
	\includegraphics[width=2.5in]{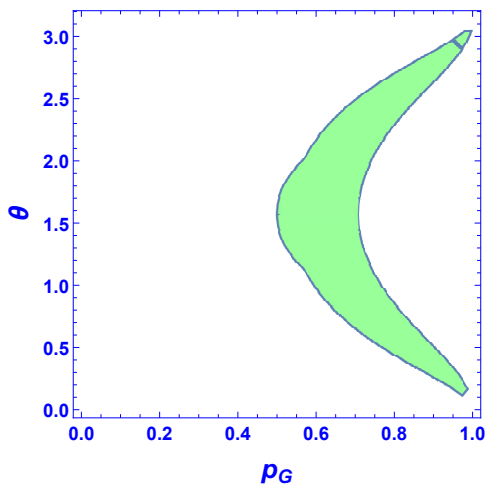}\\
	\caption{\emph{The shaded portion gives the restriction to be imposed on the noise parameter $p_G$ for which corresponding member of the noisy GHZ family(Eq.(\ref{pst8}))is conditionally steerable. However the conditional state $\rho^{BC}_{\Pi_x^{a_0}}$ is Bell-CHSH local. Clearly the restriction varies with the settings parameter $\theta$, optimal result being obtained for $\theta=\frac{\pi}{2}$(see FIG.(\ref{p12s})). }}\label{p11s}
\end{figure}

\begin{figure}[htb]
	\centering
	\includegraphics[width=2.5in]{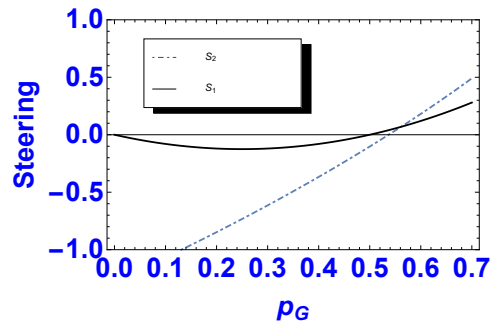}\\
	\caption{\emph{Two measures of steering $S_1$(Eq.(\ref{pst9i})) and $S_2$(Eq.(\ref{pst9ii})) are plotted against the noise parameter $p_G$ which may be considered as a measure of resistance against noise offered by the corresponding member of noisy GHZ family. The dashed curve shows variation of $S_2$ with $p_G$ whereas the solid curve shows the same for measure $S_1.$ Clearly for this class of states, $S_1$ turns out to be a better measure as conditional steerability of the state can be detected by $S_1$ for any $p_G> \frac{1}{2}$ whereas the dashed curve cuts the $p_G$ axis after $p_G=\frac{1}{2}$. }}\label{p12s}
\end{figure}

\begin{figure}[htb]
\centering
\includegraphics[width=2.5in]{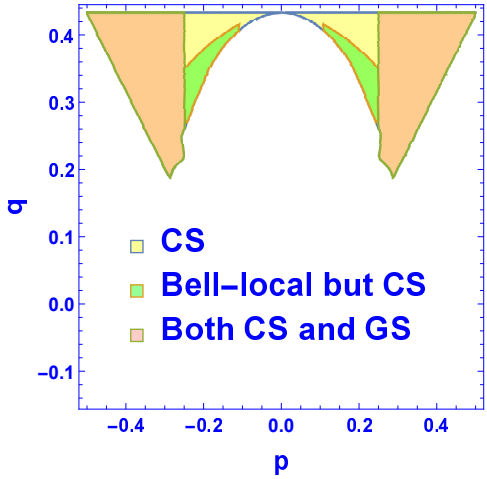}
\caption{\emph{In the figure $\textmd{CS}$ and $\textmd{GS}$ stand for `conditionally steerable' and `genuinely steerable' respectively. Clearly each of the three differently colored areas gives a restricted region in the parameter space $(p,q)$ such that any member of GHZ symmetric family characterized by parameter $p,q$ lying in any of this region is conditionally steerable. }}
\label{pic3}
\end{figure}

Let one of the three parties, Charlie(say) performs a single projective measurement on his qubit and communicates the
 output(say, $c_0$) to other two parties. Consequently other two parties Alice and Bob now share a conditional state
 $\rho^{AB}_{\Pi_z^{c_0}}$. If the state density matrix of $\rho^{AB}_{\Pi_z^{c_0}}$ satisfies atleast one of the
  sufficient criteria of steering(Eqs(\ref{st7i},\ref{st8iv})) then that guarantees that $\rho^{AB}_{\Pi_z^{c_0}}$
  is steerable from Alice to Bob. This in turn guarantees that the tripartite quantum state($\rho_{ABC}$) possesses
  conditional steering nonlocality.  However, both the steering criteria, are sufficient and not necessary.
  So if $\rho^{AB}_{\Pi_z^{c_0}}$ fails to satisfy both of them then no definite conclusion can be made. Under such
  circumstances, conditional states $\rho^{AC}_{\Pi_y^b}$
(shared between Alice and Charlie when Bob broadcasts output of his local projective measurement) and $\rho^{BC}_{\Pi_x^a}$
(shared between Bob and Charlie when Alice broadcasts output of her local projective measurement) are checked for steerability.
If atleast one of the conditional states turns out to be steerable then $\rho_{ABC}$ is conditionally steerable.
In totality if atleast one of three conditional states is steerable then the tripartite state is conditionally steerable.
However if none of them turns out to be steerable(upto $S_1$ and $S_2$), then no definite conclusion can be given about
conditional steerability of $\rho_{ABC}.$
For instance, let mixed GHZ state be shared between the three parties:
\begin{equation}\label{pst8}
\rho_{ABC} = p |GHZ^+\rangle\langle GHZ^+|+\frac{1-p}{8}\mathbb{I}_8
\end{equation}
where
\begin{equation}\label{pst9}
|GHZ^+\rangle=\frac{|000\rangle+|111\rangle}{\sqrt{2}}.
\end{equation}
and $\mathbb{I}_8$ denotes the $8\times8$ identity matrix. Let Alice performs a projective measurement in any arbitrary direction: $M_A=\overrightarrow{\alpha}.\overrightarrow{\sigma}$ where $\overrightarrow{\alpha}=(\sin\theta\cos\phi,\sin\theta\sin\phi,\cos\theta)$, $\theta,\phi\in[0,2\pi].$ Let projection of the qubit in up and down direction along the vector $\overrightarrow{\alpha}$ be denoted as $a_0$ and $a_1$ respectively. For a particular example, let $\phi=0$ and let Alice obtains output $a_0$ and broadcasts the output. Here $\rho^{BC}_{\Pi_x^{a_0}}$ denotes the conditional state. The correlation matrix $t$(Eq.(\ref{st4})) of $\rho^{BC}_{\Pi_x^{a_0}}$ has a diagonal form: $t=\textmd{diag}(p_G\sin\theta,-p_G\sin\theta,p_G)$. The elements $t_{i0}$ and $t_{0i}$ are all $0$ except $t_{30}$ and $t_{03}$ each of which turns out to be $p_G\cos\theta.$  For $\rho^{BC}_{\Pi_x^{a_0}}$, the steering criteria($S_1$, $S_2$) have the form:
\begin{equation}\label{pst9i}
S_1=\textmd{max}\{p_G,\,p_G |\sin\theta|-\frac{2}{3}p_G^2(1+2\sin^2\theta)|\}
\end{equation}
\begin{eqnarray}\nonumber
%\begin{split}
S_2=\textmd{max}\{2p_G|\sin\theta|&-&\frac{2\sqrt{1-p_G}}{\pi}
\sum_{j=0}^1\sqrt{1+p_G+p_G(-1)^j\cos\theta}
,\\ \nonumber p_G(1+\sin\theta)&-&\frac{4}{\pi}\sqrt{1-p_G^2\sin^2\theta} \}\\ \label{pst9ii}
%\end{split}
\end{eqnarray}
Again the conditional state($\rho^{BC}_{\Pi_x^{a_0}}$) is Bell-CHSH local\cite{HOR} for:
\begin{equation}\label{pst9m}
    \textmd{max}\{\sqrt{2} p_G |\sin\theta|,\,p_G\sqrt{1+\sin^2\theta}\}\leq 1.
\end{equation}
The conditional steerable states from the family(Eq.(\ref{pst8})) are those which satisfy either $S_1<0$(Eq.(\ref{pst9i}))
 or $S_2>0$(Eq.(\ref{pst9ii})) or both. So these two criteria impose some restrictions over noise parameter $p_G$ which
 varies with the measurement parameter $\theta$(see FIG.(\ref{p11s})). The optimal result is obtained for $\theta=\frac{\pi}{2}$(see FIG.(\ref{p12s})). Clearly, Eq.(\ref{pst9i}) implies that for $\theta=\frac{\pi}{2}$, $S_1<0$ for any value of $p_G >\frac{1}{2}.$ So mixed GHZ state exhibits conditional steering nonlocality for any value of the noise parameter($p_G$) greater than $\frac{1}{2}$. However, if one considers the approach of Popescu\cite{Popescu92}, i.e., considers Bell-nonlocality of the conditional states, then nonlocal behavior of mix GHZ states can only be obtained for $p_G>\frac{1}{\sqrt{2}}$(Eq.(\ref{pst9m})). \\
 After mix GHZ class of states, we now consider the GHZ symmetric class of states(Eq.(\ref{pst9iii})). GHZ symmetric states, a class of of tripartite mixed entangled states, important from quantum theoretical perspective has been topic of many recent research activities\cite{Elt1,Elt2,Elt3,Biswajit16}. The class is given by:\\
$\Phi(p, q)$$ =$$ (\frac{2q}{\sqrt{3}}+ p)|GHZ_{+}\rangle\langle GHZ_{+}| \,+$
\begin{equation}\label{pst9iii}
(\frac{2q}{\sqrt{3}}- p)|GHZ_{-}\rangle\langle GHZ_{-}| + (1- \frac{4q}{\sqrt{3}})\frac{\mathbb{I}_8}{8}
\end{equation}
where $|GHZ_{-}\rangle = \frac{|000\rangle - |111\rangle}{\sqrt{2}}$ and $|GHZ_{+}\rangle$ is given by Eq.(\ref{pst9}). The requirement $\Phi(p, q)\geq 0$ gives the constraints:
\begin{equation}\label{g1}
    -\frac{1}{4\sqrt{3}}\leq q \leq \frac{\sqrt{3}}{4}
\end{equation}
 and
\begin{equation}\label{pst9iv}
|p| \leq \frac{1}{8}+\frac{\sqrt{3}}{2} q.
\end{equation}
This family of states includes not only GHZ states but also the maximally mixed state $\frac{\mathbb{I}_8}{8}$. Let any member from this class of states be shared in between Alice, Bob and Charlie where Alice and Charlie have access to qubit measurements. Let Alice performs equatorial measurement $\frac{\sigma_z+\sigma_x}{\sqrt{2}}$. Let her qubit gets aligned along the vector $\frac{|0\rangle+|1\rangle}{\sqrt{2}}$ and she communicates the result to Bob and Charlie. The conditional state($\rho^{BC}_{\Pi_x^{a_0}}$) is steerable from Bob to Charlie if the state parameters $p,$ $q$ satisfy atleast one of the steering criteria: $S_1<0$, $S_2>0$. So conditionally steerable states from this class of states are obtained over a restricted region in state parameter space $(p,q)$(see FIG.\ref{pic3}) as detected by the two sufficient criteria of steering(Eqs.(\ref{st7i},\ref{st8iv})).\\
\\

Having introduced and citing some particular examples of conditionally steerable states from some known classes of tripartite mixed entangled states, we are now in a position to compare this weaker form of nonlocality with some existing notions of nonlocality which in turn points out that this notion of nonlocality enhances utility of any tripartite state which may otherwise be treated as useless. From application point of view, it is interesting to explore whether by using the notion of conditional steering nonlocality one can increase the resistance to noise of any noisy tripartite state. For our purpose we investigate some known families of noisy tripartite entangled states.
  \\
\subsection{Tripartite local states may be conditionally steerable}\label{tri}
\begin{figure}[b]
\centering
\includegraphics[width=2.5in]{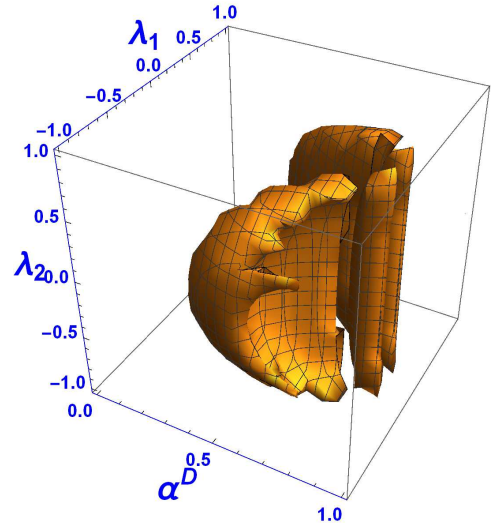}
\caption{\emph{The area gives the region in the parameter space $(\lambda_1,\lambda_2,\alpha^D)$ for which the corresponding members from the family of $\Upsilon$ are conditionally steerable and hence can be considered as a nonlocal resource. The conditional state $\Upsilon^{AC}_{\Pi_y^{b_0}}$ is however Bell-CHSH local. }}
\label{picnl1}
\end{figure}

\begin{figure}[htb]
	\centering
	\includegraphics[width=2.5in]{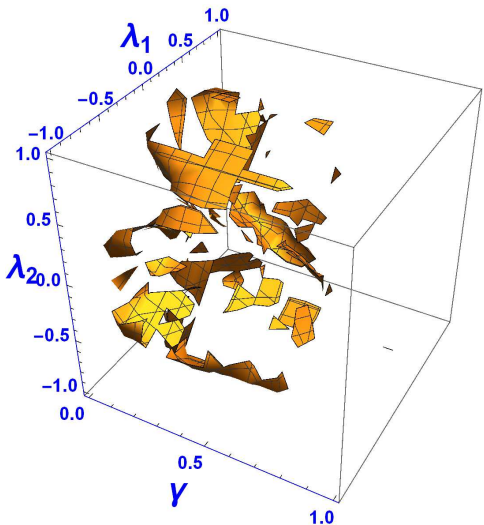}
	\caption{\emph{ Shaded region impose restrictions on the state parameters $\lambda_1$, $\lambda_2$ and noise parameter $\gamma$ such that any member from the family of $\Psi$ characterized by state parameters from this region are conditionally steerable. The conditional state $\Psi^{BC}_{\Pi_x^{a_1}}$ is however Bell-CHSH local. }}
	\label{picnl2}
\end{figure}

Here we consider the two input and two output(for each party) Bell scenario where, as already discussed before, the necessary and sufficient criteria for detecting tripartite nonlocality is given by the Bell-local polytope designed in\cite{Sliwa03}. So if any tripartite state fails to violate any of the facet inequalities  of this polytope then the state is local upto that specific scenario. However it may generate nonlocal correlations in some other measurement scenario. So in that context one may explore if one can generate nonclassical correlations from such states. Interestingly we observed that some of these local states are conditionally steerable and hence can be considered as a nonlocal resource without being subjected to any further investigation in other measurement scenarios. To be precise, we provide a few examples of mixed states which are Bell local with respect to some necessary sufficient conditions\cite{Sliwa03} but still possess conditional steering. Consider a tripartite pure state:
\begin{equation}\label{pst10}
|\psi \rangle = \lambda_0|000\rangle+\lambda_1|101\rangle+\lambda_2|110\rangle
\end{equation}
where each of $\lambda_i(i=1,2,3)\in[-1,1]$ and $\sum_{i=1}^3\lambda_i=1$. When this state is passed through a depolarization channel\cite{kr1}, the corresponding noisy state is given by:
\begin{equation}\label{pst11}
\Upsilon= \alpha^D |\psi\rangle\langle\psi|+\frac{1-\alpha^D}{8}\mathbb{I}_8.
\end{equation}
Here $\alpha^D$ denotes the visibility and hence characterizes the resistance to noise by the noisy state. Let $\Upsilon$ be shared in between Alice, Bob and Charlie. Let Bob measures his qubit in $\sigma_z$ basis and gets output $b_0$, i.e., the qubit gets aligned along up direction of $z$ axis. There exist some conditional states $\Upsilon^{AC}_{\Pi_y^{b_0}}$ that are steerable from Alice to Charlie, as detected by the steering criteria($S_1$(Eq.(\ref{st7i})),$S_2$(Eq.(\ref{st8iv}))), see FIG.\ref{picnl1}. \\
Now let us consider the noisy version of state($|\psi\rangle$) obtained after passing the state through an amplitude damping channel\cite{kr1}. Let $\Psi$ denotes the density matrix of the noisy state(see Appendix.\ref{AppA}). Let $\nu=1-\gamma$ measures the resistance to noise by $\Psi$ where the parameter $\gamma$ characterizes the amplitude damping channel. Let $\Psi$ be shared between Alice, Bob and Charlie with Alice and Charlie having access to qubit measurement. Let after measuring her qubit in $\sigma_z$ basis, Alice obtains $a_1$(along down direction of $z$- axis). The steerability of the conditional state $\Psi^{BC}_{\Pi_x^{a_1}}$  from Bob to Charlie can be detected by atleast one of the two measures of steering $S_1$(Eq.(\ref{st7i})) and $S_2$(Eq.(\ref{st8iv})) and hence $\Psi$ is conditionally steerable for some restricted range of state parameters(see FIG.\ref{picnl2}).
\\
\\
Now there exist members from each of these two families of mixed tripartite states given by the noisy versions($\Upsilon$(Eq.(\ref{pst11})), $\Psi$(Eq.(\ref{pst12}))) of the pure tripartite state $|\psi\rangle\langle\psi|$, such that those states are local in a specific Bell scenario\cite{Sliwa03} as they satisfy all $46$ facets of the Bell-local polytope\cite{Sliwa03} but are conditionally steerable and hence reveal nonlocality via the weaker notion of conditional steering nonlocality. This in turn points out the utility of this notion of nonlocality to lower down the visibility(measuring resistance to noise) of a noisy state. We enlist numerical observations in support of our claim in TABLE(1)(see Appendix.\ref{AppB}).

We have discussed in the last section about the existence of conditionally steerable states from the GHZ symmetric class of states(see FIG.3). However some of those states cannot exhibit nonlocality in this specific Bell scenario. To be specific the subclass of nonlocal states from this class are those for which the state parameters $p,q$ satisfy \cite{Biswajit16}: $L_1$$>$$2$ or $L_2$$>$$4$ or both where $L_1$ and $L_2$ are the respective bounds of violation of Mermin inequality and $15^{th}$(according to the ordering of the inequalities in \cite{Sliwa03}) facet inequality(the two most efficient detectors of standard nonlocality for this class\cite{Biswajit16}):
  \begin{equation}\label{pst12i}
  L_1 = 8|p|
  \end{equation}
  and
  \begin{equation}\label{pst12ii}
  L_2 = \max[\frac{8(9|p^3|-8\sqrt{3}|q^3|)}{9p^2-12q^2}, -16\sqrt{3}|q|].
  \end{equation}
  A comparison of these local bounds with that of two sufficient criteria of steering guarantees the existence of some conditionally steerable from this subclass which can satisfy neither $L_1$$>$$2$ nor $L_2$$>$$4$(see FIG.\ref{pic3}).
\subsection{Tripartite steering may not be necessary for conditional steering}
Comparison between tripartite nonlocality and conditional steering nonlocality, as discussed above clearly reveals the fact that a conditionally steerable state may be local in some specific  Bell scenario\cite{Sliwa03}. Intuitively such a state should not reveal any genuine form of nonlocal correlations and so they may not be even genuinely steerable\cite{Jeba}. In this context it becomes interesting to explore whether there exist states which may not reveal weaker form of steering nonlocality also but are conditionally steerable.
As we have already discussed before that there exist members from both families of noisy states $\Upsilon$ and $\Psi$ which are conditionally steerable. But none of these states can violate the steering inequality(Eq.(22) in \cite{cavals}) and hence may not be steerable in the scenario introduced in \cite{cavals}. This in turn points out the utility of conditional steerability over the stronger notion of tripartite steerability\cite{cavals}.
\subsection{Conditional steering does not require genuine entanglement}
\begin{figure}[htb]
	\centering
	\includegraphics[width=2.5in]{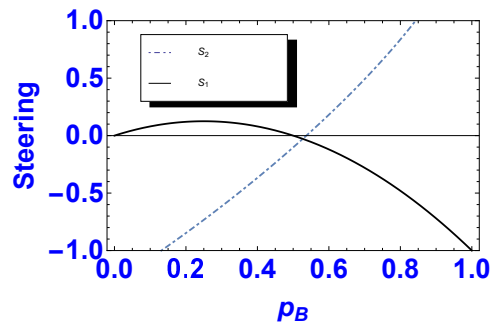}
	\caption{\emph{The range of noise parameter($p_B$) for which biseparable state(Eq.(\ref{b1})) shows conditional steerability is plotted in this figure. Clearly for this state $S_1$ serves as a better detection criterion for detecting steerability of the conditional state($\rho^{AB}_{\Pi_z^{c_1}}$). }}
	\label{picge}
\end{figure}

Till now we have discussed relation of conditional steerability of states with various existing notions of tripartite nonlocality. On the other hand entanglement is a trivial requirement for conditional steering. At this point it seems interesting to explore whether genuine entanglement is necessary for a tripartite state to be conditionally steerable. In this context it is observed that for a tripartite state to be conditionally steerable it may not be genuinely entangled. For that we have considered the class of mixed biseparable states:
\begin{equation}\label{b1}
\rho_{\textmd{bisep}}= p_B|\psi_{\textmd{bisep}}\rangle\langle\psi_{\textmd{bisep}}|+(1-p_B)\frac{\mathbb{I}}{8}, \textmd{where}
\end{equation}
$|\psi_{\textmd{bisep}}\rangle=\frac{|00\rangle+|11\rangle}{\sqrt{2}}\bigotimes\frac{|0\rangle+|1\rangle}{\sqrt{2}}.$ Let out of three parties, Bob and Charlie have access to quantum measurements. Let Charlie measures his qubit in $\sigma_z$ basis and gets output $c_1$(i.e. his qubit gets aligned along down direction of $z$ axis). The conditional state $\rho^{AB}_{\Pi_z^{c_1}}$ is steerable from Alice to Bob for some restricted range of the noise parameter $p_B$(see FIG.\ref{picge}). This in turn justifies our claim that bi-separable entanglement content of a tripartite state is sufficient for the state to become conditionally steerable. However such an observation is quite intuitive due to the fact if qubits of any two parties(say Alice and Bob) of the tripartite state are entangled, then for any arbitrarily small amount of entanglement content, the conditional bipartite state resulting due to any projective measurement by the remaining party(say Charlie) on his qubit will be entangled and hence steerable which in turn guarantees conditional steerability of the bi-separable state.
\subsection{Persistence of steering: a subclass of conditional steering}
Here we deal with an interesting implication of conditional steering nonlocality. There exist tripartite states which may fail to persist steering when subjected to particle loss, hence strong persistency of steering nonlocality\cite{per1} being $1$ but may be conditionally steerable and so having persistency of steering nonlocality greater than $1$\cite{per2,persist}.\\
 Consider a particular subclass of the mixed entangled  class of GHZ symmetric states \cite{Elt3}:
  \begin{equation}\label{pst13}
  |p|>\frac{1}{4}.
  \end{equation}
 This subclass of GHZ symmetric class of states is genuinely steerable\cite{Jeba}, i.e., violate the criteria given by Eq.(\ref{pst5}). Now each of the bipartite reduced states(obtained by tracing out one party at a time) obtained from any of these genuinely steerable states is $X$ state having density matrix with only non vanishing diagonal entries and therefore separable\cite{Swapan}. So none of the reduced states is steerable. Hence strong persistency of steering nonlocality for this class of states is $1$, i.e., minimal. \\
  Now this collection of genuinely steerable states(Eq.(\ref{pst13})) forms a subclass of conditionally steerable GHZ symmetric  states(see FIG.3). So there exist tripartite states from the GHZ symmetric class of states that are both genuinely steerable and conditionally steerable thereby indicating that persistency of steering nonlocality is $>1$.
\subsection{Conditional steerability of states not useful in teleportation}
So far we have presented foundational aspects of conditional steering. In the following part we link this scenario with an information processing task. In bipartite case there exist some states which are helpful in teleportation but may not be Bell nonlocal. In Subsection.\ref{tri} we have shown conditional steering does not necessarily imply tripartite nonlocality. So it may be an interesting question whether conditional steerable tripartite state is an useful resource for teleportation. However our answer is negative.
As has been already discussed before that a tripartite state having vanishing negativity in atleast one possible cut may not be GHZ distillable \cite{GHZ distillation} which in turn implies that those are not useful for the task of teleportation also. Here we deal with some of these types of states so as to explore whether they can reveal any weaker form of nonlocality in spite of being useless in a teleportation task. It is observed there exist some states from each of the subclasses given by $\Upsilon$ and $\Psi$ which are conditionally steerable though negativity of such  states vanish atleast in one cut(see FIG.\ref{pictele}). We enlist related numerical observations in TABLE.(2)(see Appendix.\ref{AppB}).
%\begin{widetext}
\begin{center}
\begin{figure}
\begin{tabular}{|c|c|}
\hline
\subfloat[$\Upsilon$]{\includegraphics[trim = 0mm 0mm 0mm 0mm,clip,scale=0.47]{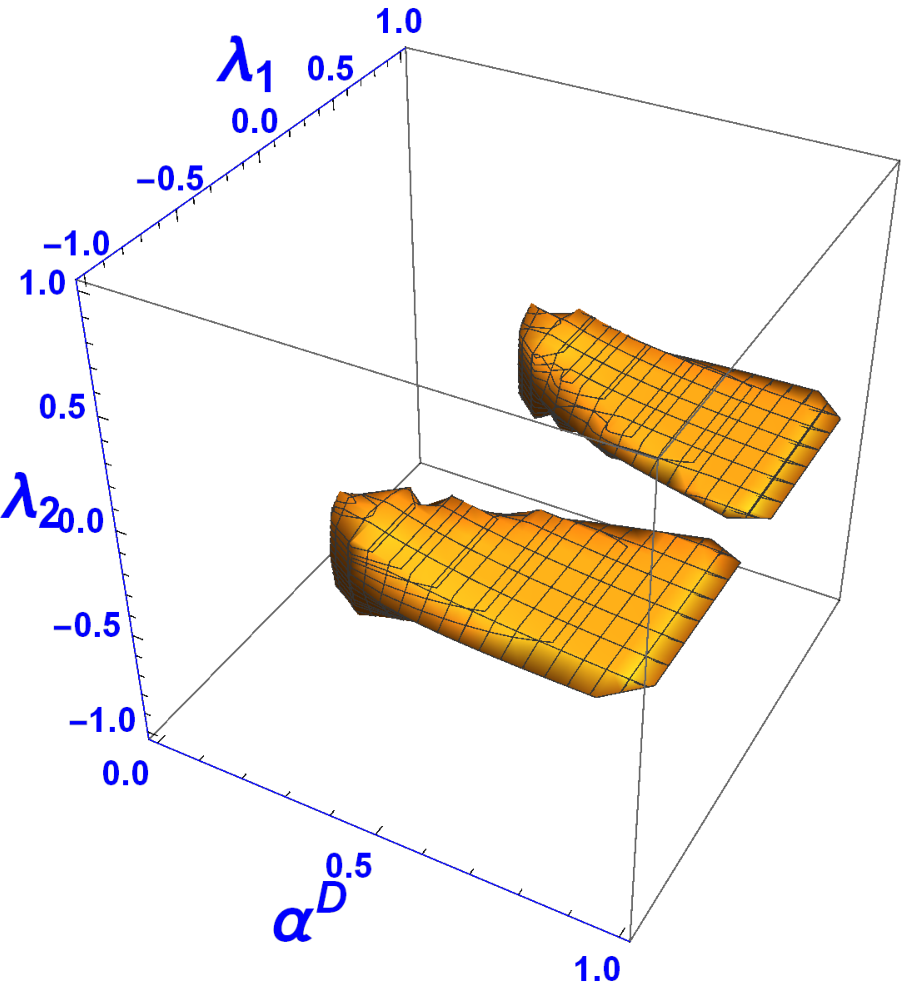}} &
\subfloat[$\Psi$]{\includegraphics[trim = 0mm 0mm 0mm 0mm,clip,scale=0.47]{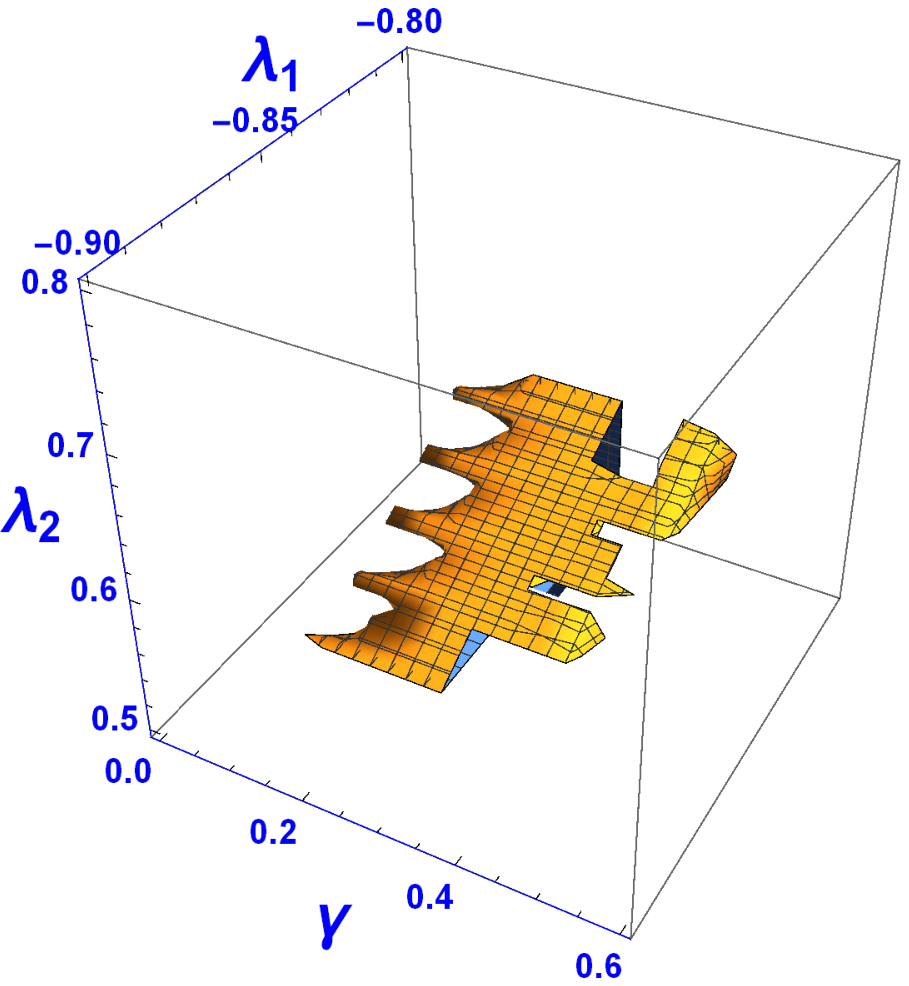}}\\
\hline
\end{tabular}
\caption{\emph{Shaded region in both the subfigures give a restricted set of state parameters $\lambda_1$ and $\lambda_2$ of the state $|\psi\rangle\langle \psi|$ given by(Eq.(\ref{pst10})) along with the noise parameter, $\alpha^D$ for depolarized version $\Upsilon$(Eq.(\ref{pst11})) of $|\psi\rangle\langle \psi|$ in sub-figure(a) and $\gamma$ for amplitude damped version of $|\psi\rangle\langle \psi|$, i.e. $\Psi$ in sub-figure(b) such that the corresponding states may not be GHZ distillable but yet they are nonlocal via the notion of conditional steering nonlocality.}}
\label{pictele}
\end{figure}
\end{center}
%\end{widetext}

\section{Discussions}\label{concl}
In \cite{Popescu92} Popescu characterized nonlocality of pure $n$-partite system by studying bipartite violation of local realism. In our present topic of discussion we have followed an analogous approach so as to explore whether such a characterization is possible for mixed entangled states also. Interestingly we have observed that such a characterization of nonlocality of mixed tripartite quantum states is possible if one considers steerability of atleast one of bipartite conditional states generated due to projective measurement done by one of the parties on his/her qubit. Any tripartite state which exhibits such weaker form of tripartite nonlocality is said to be conditionally steerable state. Such a characterization of tripartite states in turn exploits the utility of these states as a nonlocal resource which otherwise may seem to be useless if one considers the pre-existing notions of tripartite nonlocality. However there also exist some conditionally steerable states which are neither useful in tripartite teleportation task nor can be used in GHZ distillation protocol. Now the notion of conditional steering nonlocality of a tripartite state is based on some sufficient criteria for detecting bipartite steering nonlocality. So one may expect to get better result if the same can be made from some necessary and sufficient criteria of detecting bipartite steering nonlocality thereby giving a more compact characterization of tripartite steering nonlocality. However, to the best of authors' knowledge no such necessary and sufficient criteria for detecting steerability of a bipartite state exists in the literature. So it may be interesting to explore in that direction first and then redefine the concept of conditional steerability. Besides, steering phenomenon and randomness both being intrinsic features of quantum particles, it may be interesting to study the interplay between them.
\\
\emph{Acknowledgement:}
We would like to thank Prof. Guruprasad Kar for useful discussions. AM acknowledges support from the CSIR project 09/093(0148)/2012-EMR-I.

%#######################################

\appendix
\subsection{Amplitude damped noisy state}\label{AppA}
Mathematically, amplitude damping is represented using Krauss operator formalism\cite{kr1,kr2}. After passing a  qubit state $\omega$ through an amplitude damping channel, the resulting noisy state is given by:
\begin{equation}\label{r2}
    \theta_{AD}(\varrho)=F_0\omega F_0^\dag + F_1\omega F_1^\dag
\end{equation}
 where $F_0=\left(
              \begin{array}{cc}
                1 & 0 \\
                0 & \sqrt{1-\gamma} \\
              \end{array}
            \right)$
  and $F_1=\left(
              \begin{array}{cc}
                0 & \sqrt{\gamma} \\
                0 & 0 \\
              \end{array}
            \right)$ are the Krauss operators used for representing the amplitude damping operation. When each of the three qubits of a tripartite state($\rho$) is passed individually through an amplitude damping channel characterized by parameter $\gamma$, noisy version of the state is given by:
\begin{equation}\label{krauss}
    \varrho_{AMP}=P( W_i\bigotimes W_j\bigotimes W_k)\varrho P( W_i\dag\bigotimes W_j\dag\bigotimes W_k\dag)
\end{equation}
where $P(W_i\bigotimes W_j\bigotimes W_k)$ denotes all possible permutations of the Krauss operators $W_i,W_j,W_k$ over possible $i,j,k\in\{0,1\}.$ In this context,  noisy version of the state $|\psi\rangle\langle \psi|$(Eq.(\ref{pst10})) is given by:
\begin{widetext}
\begin{equation}\label{pst12}
\Psi=\left(\begin{array}{cccccccc}
\lambda_0^2+\gamma^2(\lambda_1^2+\lambda_2^2)&0&0&0&0&\nu\lambda_0\lambda_1 & \nu\lambda_0\lambda_2 \\
0&\nu\gamma\lambda_1^2 & \nu\gamma \lambda_1\lambda_2&0&0&0&0&0\\
0&\nu\gamma \lambda_1\lambda_2 & \nu\gamma\lambda_2^2&0&0&0&0&0\\
0&0&0&0&0&0&0&0\\
0&0&0&0&\nu\gamma(1-\lambda_0^2)&0&0&0\\
\nu\lambda_0\lambda_1&0&0&0&0&\nu^2\lambda_1^2&\nu^2\lambda_1\lambda_2&0\\
\nu\lambda_0\lambda_1 &0&0&0&0&\nu^2\lambda_1\lambda_2&\nu^2\lambda_2^2\\
0&0&0&0&0&0&0&0\\
\end{array} \right)
\end{equation}
where $\nu=1-\gamma$ measures the resistance to noise by $\Psi.$
\subsection{Numerical Observations}\label{AppB}
Below we present some numerical observations related to the results as discussed in the main text.
\begin{table}[htp]
  	\begin{center}
  		\begin{tabular}{|c|c|c|c|c|c|}
  			\hline
  			Channel&State Parameter& Noise Parameter&Bell nonlocal&Conditional Steerability&Range of advantage\\
  			\hline
  			Depolarization&$\lambda_1=-0.6$, $\lambda_2=0.07$& $\alpha^D$&$\alpha^D\geq 0.6594$&$\alpha^D\geq 0.359$&$\alpha^D\in[0.359,0.6594)$\\
  			\hline
  			Amplitude Damping&$\lambda_1=-0.001$, $\lambda_2=0.9011$ & $\nu$& $\nu\geq 0.70$ & $\nu\geq 0.31$& $\nu\in[0.31,0.70)$\\
  			\hline
  		\end{tabular}\\
  		\caption{The table gives example of conditionally steerable Bell-local states(for some fixed state parameters) from each of two families $\Upsilon$ and $\Psi$.}
  	\end{center}
  	\label{table1}
  \end{table}
\begin{table}[htp]
   	\begin{center}
   		\begin{tabular}{|c|c|c|c|c|c|c|}
   			\hline
   			State&State Parameter& Noise Parameter& cut in which&Negativity vanishes&Conditional&Range of\\
   			&&&negativity vanishes&&Steerability& advantage\\
   			\hline
   			$\Upsilon$&$\lambda_1=0.45$, $\lambda_2=-0.09$& $\alpha^D$&$2|13$ cut&$\alpha^D\leq0.582$&$\alpha^D\geq 0.442$&$\alpha^D\in[0.442,0.582]$\\
   			\hline
   			$\Psi$&$\lambda_1=-0.85$, $\lambda_2=0.6$ & $\nu$&$1|23$ cut&$\nu\leq 0.72$&$\nu\geq 0.55$ & $\nu\in[0.55,0.72]$\\
   			\hline
   		\end{tabular}\\
   		\caption{The table gives examples of conditionally steerable tripartite states(from each of two families $\Upsilon$ and $\Psi$) having negativity $0$ in one possible cut for some restricted range of noise parameter and hence may not be useful neither for the task of teleportation nor for the task of GHZ distillation within that range. }
   	\end{center}
   	\label{table2}
   \end{table}
   \end{widetext}


\begin{thebibliography}{99}
\bibitem{Reviw Ent} R. Horodecki, P. Horodecki, M. Horodecki, and K. Horodecki, "Quantum entanglement", \href{http://journals.aps.org/rmp/abstract/10.1103/RevModPhys.81.865}{Rev. Mod. Phys. {\bf 81}, 865 (2009)}	
\bibitem{Bennett93} C. H. Bennett, G. Brassard, C. Cr\'{e}peau, R. Jozsa, A. Peres, and W. K. Wootters, "Teleporting an unknown quantum state via dual classical and Einstein-Podolsky-Rosen channels", \href{http://dx.doi.org/10.1103/PhysRevLett.70.1895}{Phys. Rev. Lett. {\bf 70}, 1895 (1993)}.

\bibitem{Bennett92}  C. H. Bennett, S. J. Wiesner, "Communication via one- and two-particle operators on Einstein-Podolsky-Rosen states", \href{http://dx.doi.org/10.1103/PhysRevLett.69.2881}{Phys. Rev. Lett. 69, 2881 (1992)}.

\bibitem{Bell64} J. S. Bell, "On the Einstein Podolsky Rosen Paradox", Physics {\bf 1} (3): 195–200 (1964), J. S. Bell, Speakable and Unspeakable in Quantum Mechanics (Cambridge University Press, 1987).

\bibitem{random} S. Pironio, A. Ac\'{i}n, S. Massar, A. Boyer de la Giroday, D. N. Matsukevich, P. Maunz, S. Olmschenk, D. Hayes, L. Luo, T. A. Manning and C. Monroe, " Random numbers certified by Bell’s theorem", \href{http://www.nature.com/nature/journal/v464/n7291/full/nature09008.html}{Nature {\bf 464}, 1021-1024}.
R. Colbeck and R. Renner, ``Free randomness can be amplified", \href{http://www.nature.com/nphys/journal/v8/n6/full/nphys2300.html}{Nat. Phys.{\bf 8}, 450 (2012)};
\bibitem{manik}A. Chaturvedi and M. Banik, ``Measurement-device–independent randomness from local entangled states",
\href{http://iopscience.iop.org/article/10.1209/
	0295-5075/112/30003/meta;jsessionid=D6C96ABB3E61
	C42C542A9553E8A4F4DC.c3.iopscience.cld.iop.org}{EPL {\bf 112}, 30003 (2015)}.
\bibitem{key} J. Barrett, L. Hardy, and A. Kent, ``No signaling and quantum key distribution", \href{http://dx.doi.org/10.1103/PhysRevLett.95.010503}{Phys. Rev. Lett. {\bf 95}, 010503 (2005)};
 A. Ac\'{i}n, N. Gisin, and L. Masanes, ``From Bells theorem to secure quantum key distribution", \href{http://dx.doi.org/10.1103/PhysRevLett.97.120405}{Phys. Rev. Lett. {\bf 97}, 120405 (2006)};

\bibitem{dw} N. Brunner, S. Pironio, A. Ac\'{i}n, N. Gisin, A. A. Methot, and	V. Scarani, ``Testing the dimension of Hilbert spaces", \href{http://dx.doi.org/10.1103/PhysRevLett.100.210503}{Phys. Rev. Lett. {\bf 100}, 210503 (2008)};
R. Gallego, N. Brunner, C. Hadley, and A. Ac\'{i}n, ``Device independent tests of classical and quantum dimensions",  \href{http://dx.doi.org/10.1103/PhysRevLett.105.230501}{Phys. Rev. Lett. {\bf 105}, 230501 (2010)};
S. Das, M. Banik, A. Rai, MD R. Gazi, and S.Kunkri, ``Hardy's nonlocality argument as a witness for postquantum correlations",
\href{https://journals.aps.org/pra/abstract/10.1103/PhysRevA.87.012112}{Phys. Rev. A {\bf 87}, 012112 (2013)};
A. Mukherjee, A. Roy, S. S. Bhattacharya, S. Das, Md. R. Gazi, and M. Banik, ``Hardy's test as a device-independent dimension witness",
\href{https://journals.aps.org/pra/abstract/10.1103/PhysRevA.92.022302}{Phys. Rev. A {\bf 92}, 022302 (2015)};

\bibitem{game} N. Brunner and N. Linden, ``Connection between Bell nonlocality and Bayesian game theory", \href{http://www.nature.com/ncomms/2013/130703/ncomms3057/full/ncomms3057.html#references}{Nature Communications {\bf 4}, 2057 (2013)}.
A. Pappa \emph{et al.} ``Nonlocality and Conflicting Interest Games",
\href{http://journals.aps.org/prl/abstract/10.1103/PhysRevLett.114.020401}{Phys. Rev. Lett. {\bf 114}, 020401 (2015)}.
A. Roy, A. Mukherjee, T. Guha, S. Ghosh, S. S. Bhattacharya, M. Banik, ``Nonlocal correlations: Fair and Unfair Strategies in Bayesian Game'', \href{http://arxiv.org/abs/1601.02349}{Arxiv: 1601.02349}.
\bibitem{km1} Mukherjee, K., Paul, B. and Sarkar,D, `` Correlations in n-local scenario'',\href{http:// Quantum Inf Process. /14/ 2025}{ Quantum Inf Process. 14,2025, (2015)}.
\bibitem{Wiseman07} H. M. Wiseman, S. J. Jones, and A. C. Doherty, ``Steering, Entanglement, Nonlocality, and the Einstein-Podolsky-Rosen Paradox'', \href{http://journals.aps.org/prl/abstract/10.1103/PhysRevLett.98.140402}{Phys. Rev. Lett. 98, 140402}, S. J. Jones, H. M. Wiseman, and A. C. Doherty, ``Entanglement, Einstein-Podolsky-Rosen correlations, Bell nonlocality, and steering'', \href{http://journals.aps.org/pra/abstract/10.1103/PhysRevA.76.052116}{Phys. Rev. A 76, 052116 (2007)}
\bibitem{ban} J.-D. Bancal, J. Barrett, N. Gisin, and S. Pironio, ``The definition Of Multipartite Nonlocality'',\href{http:// Phys. Rev.A .88.014102}{Phys. Rev. A 88, 014102, 2013}.
\bibitem{kmb} Mukherjee, K., Paul, B., and Sarkar, D,`` Efficient test to demonstrate genuine three particle nonlocality'',\href{http:// J. Phys. A: Math. Theor./48/465302}{J.Phys. A 48, 465302 (2015)}.
\bibitem{Gisin91} N. Gisin, ``Bell's inequality holds for all non-product states'', \href{http://www.sciencedirect.com/science/article/pii/037596019190805I}{Phys. Lett A {\bf 154}, 201 (1991)}.

\bibitem{Gisin92} N. Gisin, A. Peres, ``Maximal violation of Bell's inequality for arbitrarily large spin'', \href{http://www.sciencedirect.com/science/article/pii/037596019290949M}{Phys. Lett. A {\bf 162}, 15 (1992)}.

\bibitem{Popescu92} S. Popescu, D. Rohrlich, ``Generic quantum nonlocality'', \href{http://www.sciencedirect.com/science/article/pii/037596019290711T}{Phys. Lett. A {\bf 166}, 293 (1992)}.
\bibitem{Sliwa03} C. Sliwa, ``Symmetries of the Bell correlation inequalities'', \href{http://www.sciencedirect.com/science/article/pii/S0375960103011150}{Phys. Lett. A, {\bf 317}, 165 (2003)}.
\bibitem{cavals}E. G. Cavalcanti, Q. Y. He, M. D. Reid, and H. M. Wiseman, ``Unified criteria for multipartite quantum nonlocality",
\href{https:http://journals.aps.org/pra/abstract/10.1103/PhysRevA.84.032115}{Phys. Rev. A {\bf 84}, 032115 (2011)};
\bibitem{Bowles14} J. Bowles, T. V\'{e}rtesi, M. T. Quintino, and N. Brunner, ``One-way Einstein-Podolsky-Rosen Steering'', \href{http://dx.doi.org/10.1103/PhysRevLett.112.200402}{Phys. Rev. Lett. {\bf 112}, 200402 (2014)}.

\bibitem{Jevtic15} S. Jevtic, M. J. W. Hall, M. R. Anderson, M. Zwierz, and H. M. Wiseman, ``Einstein–Podolsky–Rosen steering and the steering ellipsoid'', \href{https://www.osapublishing.org/josab/abstract.cfm?uri=josab-32-4-A40}{Journal of the Optical Society of America B, {\bf 32}, A40 (2015)}.

\bibitem{Jevtic16} S. J. Jones , H. M. Wiseman, ``Nonlocality of a single photon: Paths to an Einstein-Podolsky-Rosen-steering experiment'', \href{http://dx.doi.org/10.1103/PhysRevA.84.012110}{PhysRevA.\textbf{84}.012110}.
\bibitem{Zukowski} M. Zukowski, A. Dutta, and Z. Yin, ``Geometric Bell-like inequalities for steering'', \href{http://dx.doi.org/10.1103/PhysRevA.91.032107}{Phys. Rev. A {\bf 91}, 032107 (2015)}.


\bibitem{Zyczkowski} K. Zyczkowski, P. Horodecki, A. Sampera and M. Lewenstein, \href{http://dx.doi.org/10.1103/PhysRevA.58.883}{ Phys. Rev. A \textbf{58}, 883-892 (1998).}
\bibitem{vidal} G. Vidal and R. F. Werner, \href{http://dx.doi.org/10.1103/PhysRevA.65.032314}{Phys. Rev. A  \textbf{65}, 032314 (2002).}
    \bibitem{peres} A.Peres, Phys. Rev. Lett.  \textbf{76}, 1413-1416 (1996).
\bibitem{GHZ distillation} S. Lee, J. Joo, and J. Kim, \href{http://dx.doi.org/10.1103/PhysRevA.76.012311}{ Phys. Rev. A \textbf{76}, 012311 (2007).}
\bibitem{HOR} R. Horodecki, P. Horodecki, M. Horodecki,``Violating Bell inequality by mixed spin-$\frac{1}{2}$ states: Necessary and sufficient condition''\href{http://www.sciencedirect.com/science/article/pii/037596019500214N}{ Phys.Lett. A {\bf 200},340 (1995)}
\bibitem{Elt1} C. Eltschka and J. Siewert,\href{http://journals.aps.org/prl/abstract/10.1103/PhysRevLett.108.020502} Phys. Rev. Lett. \textbf{108}, 020502,(2012).
\bibitem{Elt2} J. Siewert and C. Eltschka,\href{http://journals.aps.org/prl/abstract/10.1103/PhysRevLett.108.230502} Phys. Rev. Lett. \textbf{108}, 230502,(2012).
\bibitem{Elt3} C. Eltschka and J. Siewert, Quant. Inf. Comp. \textbf{13}, 210,(2013).
\bibitem{Biswajit16} B. Paul, K. Mukherjee, D. Sarkar, ``Nonlocality of three-qubit Greenberger-Horne-Zeilinger-symmetric states'', \href{http://dx.doi.org/10.1103/PhysRevA.94.032101}{Phys. Rev. A {\bf 94}, 032101 (2016)}.
\bibitem{kr1} M. A. Nielsen and I. L. Chuang, ”Quantum computation and quantum information”, Cambridge University Press
(2000).
\bibitem{Jeba} C. Jebaratnam, ``Detecting genuine multipartite entanglement in steering scenarios'', \href{http://journals.aps.org/pra/abstract/10.1103/PhysRevA.93.052311}{Phys. Rev. A {\bf 93}, 052311 (2016)}
\bibitem{SVE} G. Svetlichny, Phys. Rev. D \textbf{35}, 3066 (1987).
\bibitem{per1} N. Brunner and T. Vertesi, ``Persistency of entanglement
and nonlocality in multipartite quantum systems'', \href{http://journals.aps.org/pra/abstract/10.1103/PhysRevA.86.042113} {Phys.
Rev. A {\bf 86}, 042113 (2012)}
\bibitem{per2} H. J. Briegel and R. Raussendorf, ``Persistent Entanglement
in Arrays of Interacting Particles'', \href{http://journals.aps.org/prl/abstract/10.1103/PhysRevLett.86.910}Phys. Rev. Lett. {\bf86},
910(2001).
\bibitem{persist}B. Paul, K. Mukherjee, A. Sen, D. Sarkar, A. Mukherjee, A. Roy, S. Bhattacharya ``Persistency of Genuine Correlations Under Particle Loss'',
\href{http://arxiv.org/abs/1609.02969v1}{Arxiv: 1609.02969v1};
\bibitem{Swapan} S. Rana, P. Parashar, ``Maximally discordant separable two-qubit \(X\) states'', \href{http://link.springer.com/article/10.1007/s11128-014-0865-0}{Quantum Inf. Process (2014) {\bf 13}: 2815 }.
\bibitem{kr2}  K. Kraus, States, Effects and operations (Springer-Verlag, Berlin, 1983).

\end{thebibliography}
\end{document}